\documentclass[epj,nopacs]{svjour}
\usepackage{dcolumn}

\usepackage{dcolumn}
\usepackage{graphicx}
\usepackage{amsmath}
\usepackage{amssymb}
\usepackage{bm}
\usepackage{maybemath}
\usepackage{revsymb}

\newcommand{\dd}{{\mathrm d}}
\newcommand{\ee}{{\mathrm e}}
\newcommand{\xmax}{x_{\mathrm max}}

\journalname{EPJ C}

\begin{document}

\setcounter{page}{1}
\title{
Revisiting unitarity corrections for electromagnetic
processes in collisions of relativistic nuclei}
\titlerunning{Revisiting unitarity corrections for electromagnetic
processes in collisions of relativistic nuclei}
\authorrunning{U. D.~Jentschura et al.}
\author{U.~D.~Jentschura 
\inst{1,2}
\and K.~Hencken \inst{3,4}
\and V.~G.~Serbo \inst{2,5,}\thanks{Corresponding author: serbo@math.nsc.ru}}

\institute{{\addrHDMPI} \and {\addrHDUNI} \and
{\addrIPBASEL} \and {\addrABBBaden} \and
{\addrNOVOSIBIRSKST}}

\date{Received: }

\newcommand{\addrHDUNI}{Institut f\"ur Theoretische Physik,
Philosophenweg 16, 69120 Heidelberg, Germany}

\newcommand{\addrNOVOSIBIRSKST}{Novosibirsk State University, Pirogova 2, 630090
Novosibirsk, Russia}

\newcommand{\addrHDMPI}{Max-Planck-Institut f\"ur
Kernphysik, Saupfercheckweg 1, 69117 Heidelberg, Germany}

\newcommand{\addrIPBASEL}{Institut f\"{u}r Physik,
Universit\"at Basel, Klingelbergstr. 82, 4056 Basel,
Switzerland}

\newcommand{\addrABBBaden}{ABB Switzerland Ltd., Corporate Research,
Segelhof 1K, 5405 Baden, Switzerland}

\abstract{ Unitarity corrections to several electromagnetic
processes in collisions of relativistic heavy nuclei are
considered. They are due to the unitarity requirement for
the $S$-matrix and correspond to the exchange of
light-by-light scattering block between colliding nuclei.
We obtain improved results for the corrections to $e^+
\,e^-$ and $\mu^+ \mu^-$ pair production, as well as new
results for unitarity corrections to the production of
photons via virtual Compton and virtual Delbr\"{u}ck
scattering. These corrections can be numerically large;
e.g., the $\mu^+ \mu^-$ pair production cross section is
reduced by about 50 \% and the nuclear bremsstrahlung by
about $15\div 20$~\%.}


\maketitle

%
%
\section{Introduction}

The subject of this paper are so-called unitarity
corrections, which form a conceptually interesting class of
corrections for quantum electrodynamic (QED) processes. The
unitarity corrections come from the unitarity requirement
for the $S$-matrix and are relevant for processes in which
the lowest order amplitude is large, i.e., in which a large
number of photons and/or real electron-positron pairs are
typically produced. One class of such processes is
bremsstrahlung and lepton pair production in
ultra-relativistic heavy-ion collisions.

In order to put the current investigation into perspective,
let us briefly recall that heavy-ion collisions definitely
concern matter under extreme conditions. The impact
parameters $\rho$ relevant for all cross sections under
study in the current paper are in the range of the electron
Compton wavelength,
\begin{equation}
\rho \sim \lambdabar_e = \frac{\hbar}{m c} = 386 \, {\mathrm{fm}} \,.
\end{equation}
Now, the typical electric fields, as seen in the laboratory frame,
generated by a particle moving at a speed characterized by a
relativistic factor $\gamma$, are of the order of
\begin{eqnarray}
{\mathcal E}(\gamma) &=& \frac{Z\,e}{\lambdabar_e^2} \,
\gamma = \frac{Z\,e\,\gamma\,m^2\,c^2}{\hbar^2}=
 \nonumber\\
 &=& \frac{m^2
\, c^3}{\hbar\,e} \, \frac{Z\,e^2}{\hbar\,c}\,\gamma =
{\mathcal E}_{\mathrm{cr}} \, Z\alpha \,\gamma \,,
\end{eqnarray}
where ${\mathcal E}_{\mathrm{cr}}$ is Schwinger's critical
field strength. The quantity $Z\alpha \,\gamma$ assumes
values in the range of $\sim 60$ for RHIC and $\sim 1800$
for the LHC, illustrating that the flash field accompanying
the nuclei may well exceed the critical field by several
orders of magnitude, for very small spatial regions and
very small times.

In view of the huge pair production and photoproduction
cross sections encountered under these conditions, it is
natural that the importance of unitarity corrections has
been recognized for the first time within the production of
electron-positron pairs in collisions of heavy nuclei (see
Ref.~\cite{Baur} and the
reviews~\cite{BaEtAl2002,BaEtAl2007}). The unitarity
correction for the one electron-positron pair production
process
\begin{equation}
Z_1+Z_2\to Z_1+Z_2+e^+e^-
\label{eepair}
\end{equation}
has been calculated in Ref.~\cite{LeMiSe2002} and found
to be about $3\div 4$~\%. Estimates of unitarity
corrections for the $\mu^+\mu^-$ single-pair production process
\begin{equation}
Z_1+Z_2\to Z_1+Z_2+\mu^+\mu^-
\label{mumupair}
\end{equation}
have been obtained in Ref.~\cite{HeKuSe2007}; in this case
the unitarity correction is found to be large ($\sim 50$~\%).

\begin{figure}[htb]
\begin{center}
\begin{minipage}{1.0\linewidth}
\begin{center}
\includegraphics[width=0.6\linewidth]{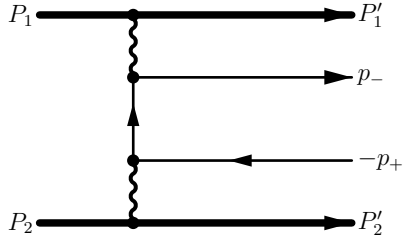}
\caption{\label{fig1} Feynman diagram for $e^+e^-$ pair
production in heavy-ion collision (first Born
approximation). {\it Bold lines} denote nuclei, {\it thin
lines} denote electrons.}
\end{center}
\end{minipage}
\end{center}

\end{figure}

\begin{figure}[htb]
\begin{center}
\begin{minipage}{1.0\linewidth}
\begin{center}
\includegraphics[width=1.0\linewidth]{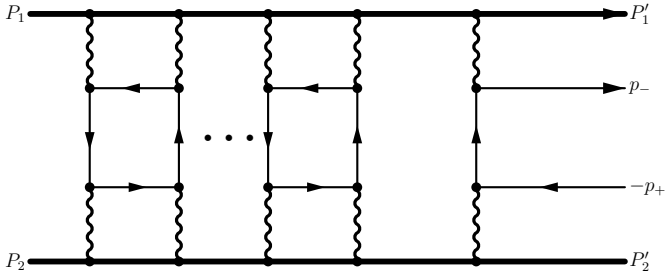}
\caption{\label{fig2} Feynman diagram for the unitarity
correction to $e^+e^-$ pair production in heavy-ion
collisions.}
\end{center}
\end{minipage}
\end{center}
 \end{figure}

Let us consider these corrections conceptually using the
process (\ref{eepair}) as an example. In this case the
lowest order Feynman diagram is represented by
Fig.~\ref{fig1}, while the diagrams of the type
depicted in Fig.~\ref{fig2}
correspond to the unitarity correction. These
diagrams include blocks of virtual light-by-light
scattering via an electron loop, whose imaginary part
corresponds to the production of electron-positron pairs by
the Cutkosky rules.

%
%

%
%

For $Z_1 \, Z_2 \, \alpha\gg 1$ and $\gamma\gg 1$, it is possible
to treat the nuclei as sources of an external field and to
calculate the probability of $n$-pair production
$P_n(\rho)$ in collisions of two nuclei at a fixed impact
parameter~$\rho$ \cite{BaEtAl2003}. The sum over $n$ of the
probabilities $P_n(\rho)$ for $n$-pair production
must be unity. The unitarity requirement is fulfilled
by the Poisson distribution
\begin{equation}
\label{poisson}
P_n(\rho)= \frac{{\bar n}_e^n}{n!} \, {\rm e}^{-{\bar n}_e}\,,
\end{equation}
whose sum over $n$ gives one. Here ${\bar n}_e\equiv{\bar
n}_e(\rho)$ is the averaged number of produced pairs at
a given $\rho$, and the factor ${\rm exp}(-{\bar n}_e)$ is the
vacuum-to-vacuum transition probability
\begin{equation}
P_0(\rho)={\rm e}^{-{\bar n}_e}=1-\sum_{n=1}^{\infty}\,P_n(\rho)\,.
\end{equation}
Roughly speaking, the probability for producing one pair,
given in perturbation theory by ${\bar n}_e$, should be
modified to read ${\bar n}_e\, \exp(-{\bar n}_e)$, and this
correction is not small for an appreciable value of ${\bar
n}_e$. This means that also the cross section for the
one-pair production $\sigma_{1}$ should be multiplied by an
appropriate factor $\exp(-{\bar n}_e)$ in the integral over
the impact parameter, which corresponds to the following
replacement,
\begin{equation}
\label{unit1ee}
\sigma_{1} = \int  {\bar n}_e(\rho)\,\dd^2{\rho} 
\; \to \; \sigma_{1} + \sigma_{1}^{\rm unit} = 
\int {\bar n}_e(\rho)\, 
{\rm e}^{-\bar{n}_e(\rho)}\,
\dd^2{\rho}\,,
\end{equation}
where, finally,
\begin{equation}
\label{unit2ee}
\sigma_{1}^{\rm unit} = - \int {\bar n}_e(\rho)\, \left[ 1
- {\rm e}^{-\bar{n}_e(\rho)} \right]\,\dd^2{\rho}
\end{equation}
is the unitarity correction to the one pair production
cross section.

In Refs.~\cite{LeMiSe2002} and \cite{HeKuSe2007},
rather rough approximations were used for the function
$\bar{n}_e(\rho)$. Here, our intention is to use
improved approximations for these functions. Thus, the aims of the present
paper are: (i) to revise the problem of the
impact-parameter dependent pair production probability
$\bar{n}_e(\rho)$, (ii) to update the unitarity corrections for the
previously calculated processes (\ref{eepair}) and
(\ref{mumupair}), (iii) to calculate the unitarity correction
for photon emission at nuclear collisions
\begin{equation}
Z_1+Z_2 \to Z_1+Z_2+\gamma\,.
\label{zzgamma}
\end{equation}

In the latter case, the probability for the production of one
and only one photon in a heavy-ion collision is modified by
the necessity of suppressing the possibility of
simultaneous production of electron-positron pairs. In the
impact-parameter representation, this implies that the
cross section for photo-production $\sigma_{\gamma}$ should
also be multiplied by the factor $\exp(-{\bar n}_e)$. This
corresponds to the replacement
\begin{equation}
\label{unit1gammaD} 
\sigma_{\gamma } = 
\int \,P_{\gamma}(\rho)\,\dd^2{\rho} 
\; \to \;
\sigma_{\gamma} + \sigma^{\rm unit}_{\gamma} = 
\int \,P_{\gamma}(\rho)\, {\rm e}^{-\bar{n}_e(\rho)}\,
\dd^2{\rho}\,,
\end{equation}
where $P_{\gamma}(\rho)$ is the probability to emit a photon
in the collision of two nuclei at a given impact parameter
$\rho$. Therefore, the unitarity correction in this case is
given by the expression
\begin{equation}
\label{unit2gammaD} \sigma^{\rm unit}_{\gamma} = - \int \,
P_{\gamma}(\rho)\, \left[1 - {\rm
e}^{-\bar{n}_e(\rho)}\right]\,\dd^2{\rho}\,.
\end{equation}

\begin{figure}[htb]
\begin{center}
\begin{minipage}{1.0\linewidth}
\begin{center}
\includegraphics[width=1.0\linewidth]{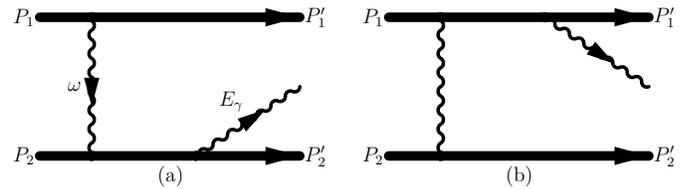}
\caption{\label{fig3} Ordinary nuclear bremsstrahlung is
the emission of a photon in a nuclear collision via a
virtual Compton subprocess.}
\end{center}
\end{minipage}
\end{center}
\end{figure}

\begin{figure}[htb]
\begin{center}
\begin{minipage}{1.0\linewidth}
\begin{center}
\includegraphics[width=1.0\linewidth]{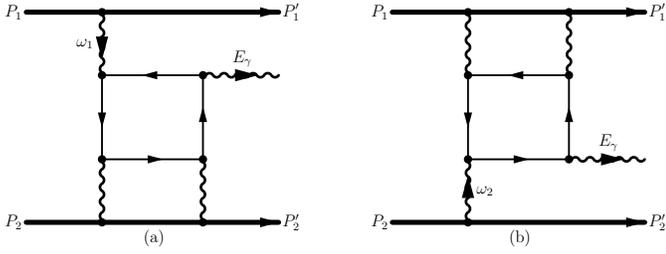}
\caption{\label{fig4} Emission of a photon in a nuclear
collision via virtual Delbr\"uck scattering in the lowest
order of QED.}
\end{center}
\end{minipage}
\end{center}
\end{figure}

It should be mentioned that there are two different
mechanisms for the photon emission in nuclear collisions:
the ordinary bremsstrahlung via virtual Compton
scattering (see Fig.~\ref{fig3}) and the recently considered
(see Refs.~\cite{GiJeSe2008plb,GiJeSe2008epjc}) emission of
photon via virtual Delbr\"uck scattering as illustrated in Fig.~\ref{fig4}.

%
%

Both of the above mentioned corrections to $e^+ e^-$ pair
production and to photoproduction correspond to the
exchange of virtual light-by-light scattering interactions
between the nuclei. Each block of light-by-light scattering
brings in an additional factor $(Z_1\alpha\, Z_2\alpha)^2$
in the amplitude of the corresponding process, and
therefore, such corrections can be omitted for the
scattering of light ions, for muon-nucleus or
electron-nucleus scattering. We definitely need $Z$ to be
large for the correction to be appreciable, and
consequently this paper is focused on heavy-ion collisions.

We organize our paper as follows. In Sec.~\ref{relphys}, we
briefly recall relevant physical parameters for modern
heavy-ion machines and discuss new approximation for the
function $\bar{n}_e(\rho)$. Section~\ref{calc} is devoted
to the actual calculation of the corrections. Specifically,
we consider unitarity corrections to $e^+ \,e^-$ and $\mu^+
\mu^-$ pair production in Secs.~\ref{calcee}
and~\ref{calcmumu}, respectively, and we reserve the
calculation of unitarity corrections to the production of
photons via virtual Compton scattering to
Sec.~\ref{calccomp} and via virtual Delbr\"{u}ck scattering
to Sec.~\ref{calcdel}. Finally, some conclusions are drawn
in Sec.~\ref{conclu}.

%
%
\section{Toward a Revised Representation of the Impact-Parameter Dependent
Pair Production Probability}
\label{relphys}

Recently, electromagnetic processes in ultra-relativistic
nuclear collisions have found strong and partially renewed
interest in numerous papers (see the reviews~\cite{BaEtAl2002,BaEtAl2007}
and references therein). Of topical importance are the RHIC
collider and the future LHC Pb--Pb option. It is therefore
useful to recall the basic physical parameters of these
colliders, namely the charge numbers of nuclei
$Z_1=Z_2\equiv Z$ and their Lorentz factors
$\gamma_1=\gamma_2\equiv \gamma$. These are given in
Table~\ref{t1}, which is cited here from
Ref.~\cite{YaEtAl2006}. Many of electromagnetic
particle-production processes are of imminent importance
for two reasons: they are either ``dangerous,'' e.g.~in
terms of possible beam losses and background, or they are
by contrast quite useful for monitoring some experiments at
the RHIC and LHC colliders \cite{BaEtAl2008}.

\begin{table}[ht]
\begin{center}
\begin{minipage}{8.0cm}
\begin{center}
\vspace{5mm} {\renewcommand{\arraystretch}{1.5}
\caption{\label{table1} Nuclear charge numbers $Z$ and
relativistic $\gamma$ factors for modern heavy-ion
machines}
\begin{center}
\par
\begin{tabular}{c@{\hspace{0.7cm}}c@{\hspace{0.4cm}}c}%
\hline
Collider & $Z$ & $\gamma$ \\
\hline RHIC, Au--Au & 79 & 108 
\\ \hline
LHC, Pb--Pb & 82 & 3000 
\\
\hline
\end{tabular}
\label{t1}
\end{center}
}
\end{center}
\end{minipage}
\end{center}
\end{table}

To fix the conventions used, we mention that natural units
with $\hbar = c = 1$ and with the fine-structure constant
$\alpha\approx 1/137$ are used throughout the text, and we
denote the electron and muon masses by $m$ and $\mu$,
respectively.

As it was mentioned in the introduction, the importance of
unitarity corrections has been recognized for the first
time at the process of electron-positron pair production
because a large number of real electron-positron pairs are
typically produced in ultra-relativistic heavy-ion
collisions. In the lowest QED order (Born approximation)
this process is described by the Feynman diagram of
Fig.~\ref{fig1}. Let ${\bar n}_e \equiv{\bar n}_e(\rho)$ be
the expected (average) number of pairs to be produced in
the collision of two nuclei at a given impact parameter
$\rho$. A closed form of the corresponding expression was
obtained in Refs.~\cite{SeWe1998,BaLe1998} 
although a complete and consistent interpretation of the
expressions found was only given later in
Ref.~\cite{BaGeLePe2001}. One issue is that ${\bar
n}_e(\rho)$ derived in
Refs.~\cite{SeWe1998,BaLe1998} 
actually requires a further regularization, which was
implemented in Refs.~\cite{LeMi2000,LeMi2001}.

As evident from Eqs.~(\ref{unit2ee}) and
(\ref{unit2gammaD}), the function ${\bar n}_e(\rho)$ is a
very important quantity for the evaluation of unitarity
corrections. However, the obtained close form for
$\bar{n}_e(\rho)$ is, in fact, a nine-fold integral and its
calculation is very laborious. Therefore, a simpler
approximate expression for $\bar{n}_e(\rho)$ is very
desirable.

The properties of  $\bar{n}_e(\rho)$ have been studied in
detail in Refs.~\cite{LeMiSe2002,LeMi2007}. The
corresponding functional form for identical heavy nuclei in
the region $\rho< \gamma/m$ reads
 \begin{eqnarray}
 \label{milsteinparam}
{\bar n}_e(\rho,\gamma,Z)&=&(Z\alpha)^4 \,F(x,Z)\,
\left[L-G(x,Z)\right]\,,
 \\
L &=& \ln{(\gamma^2)}\,, \quad x = m \, \rho\,.
 \nonumber
 \end{eqnarray}
Here, $\gamma\gg 1$ is the usual relativistic factor (see
Table 1).  The analytical expressions for the functions
$F(x,Z)$ and $G(x,Z)$ have been obtained in
Refs.~\cite{LeMiSe2002,LeMi2007} only for large values of
the impact parameters, $1\ll x = m \, \rho < \gamma$ (see
the Appendix):
 \begin{eqnarray}
 \label{bethemaximon}
F(x,Z) &=&  \frac{56}{9 \, \pi^2} \, \frac{\ln{x} -
f(Z\alpha)}{x^2}\,,
 \\
F(x,Z) \cdot G(x,Z) &=& \frac{56}{9 \, \pi^2} \,
\frac{(3/2)\ln{x}- f(Z\alpha)}{x^2}\,\ln{x}\,,
 \nonumber
 \end{eqnarray}
where
\begin{equation}
f(Z\alpha) = (Z\alpha)^2 \, \sum_{n=1}^{\infty}
\frac{1}{n[n^2+(Z\alpha)^2]}
\end{equation}
is the well-known Bethe-Maximon function
[in particular, $f(79\, \alpha) = 0.3129\,,\;
f(82 \, \alpha) = 0.3318$]. On the other
hand, for calculations related to unitarity corrections we
should know the functions $F(x,Z)$ and $G(x,Z)$ in the
range $x\sim 1$.

In Ref.~\cite{LeMi2007} a closed-form expression
for the function $F(x,Z)$ at intermediate impact parameters
in the form of five-fold integral has been given. Tables provided in
Ref.~\cite{LeMi2007} give a very clear numerical picture of
the function $F(x,Z)$ for $x=0.01\div 100$ and several
important values of $Z$. For the function $G(x,Z)$, the
approximation
\begin{equation}
G(x,Z)\approx 1.5\,\ln{(x+1)} + 1.2
\label{GofLM}
\end{equation}
(independent of $Z$) has been given in Ref.~\cite{LeMi2007}
as a rough indicator of the nonlogarithmic (in $\gamma$)
term in Eq.~(\ref{milsteinparam}).

\begin{table*}
\begin{center}
\begin{minipage}{14cm}
\begin{center}
\caption{\label{table2} Functions $A(x)$ and $B(x)$ as
given Eq.~\eqref{henckenparam}, calculated using the
approach outlined previously in Ref.~\cite{HeTaBa1994}}
\begin{tabular}{r@{\hspace{0.0cm}}lr@{\hspace{0.0cm}}lr@{\hspace{0.0cm}}l%
r@{\hspace{0.0cm}}lr@{\hspace{0.0cm}}lr@{\hspace{0.0cm}}l%
r@{\hspace{0.0cm}}lr@{\hspace{0.0cm}}lr@{\hspace{0.0cm}}l}
\hline
\multicolumn{2}{c}{\rule[-2mm]{0mm}{6mm} $x$} &
\multicolumn{2}{c}{$A(x)$} &
\multicolumn{2}{c}{$B(x)$} &
\multicolumn{2}{c}{$x$} &
\multicolumn{2}{c}{$A(x)$} &
\multicolumn{2}{c}{$B(x)$} &
\multicolumn{2}{c}{$x$} &
\multicolumn{2}{c}{$A(x)$} &
\multicolumn{2}{c}{$B(x)$} \\
\hline
%
0.&01 & 3.&537 &   8.&200 &
0.&40 & 1.&028 &   2.&910 &
4.&50 & 0.&07273 & 0.&3355 \\
0.&02 & 3.&082 &   7.&434 &
0.&50 & 0.&8879 &  2.&573 &
5.&00 & 0.&06191 & 0.&2956 \\
0.&03 & 2.&805 &   6.&927 &
0.&60 & 0.&7767 &  2.&285 &
5.&50 & 0.&05327 & 0.&2619 \\
0.&04 & 2.&611 &   6.&535 &
0.&70 & 0.&6895 &  2.&085 &
6.&00 & 0.&04635 & 0.&2339 \\
0.&05 & 2.&460 &   6.&199 &
0.&90 & 0.&5557 &  1.&747 &
6.&50 & 0.&04072 & 0.&2104 \\
0.&06 & 2.&331 &   5.&892 &
1.&00 & 0.&5039 &  1.&614 &
7.&00 & 0.&03606 & 0.&1903 \\
0.&07 & 2.&218 &   5.&613 &
1.&50 & 0.&3293 &  1.&142 &
7.&50 & 0.&03217 & 0.&1730 \\
0.&08 & 2.&122 &   5.&366 &
2.&00 & 0.&2323 &  0.&8581 &
8.&00 & 0.&02887 & 0.&1579 \\
0.&09 & 2.&037 &   5.&139 &
2.&50 & 0.&1725 &  0.&6716 &
8.&50 & 0.&02604 & 0.&1446 \\
0.&10 & 1.&962 &   4.&949 &
3.&00 & 0.&1331 &  0.&5431 &
9.&00 & 0.&02360 & 0.&1329 \\
0.&20 & 1.&486 &   3.&900 &
3.&50 & 0.&1066 &  0.&4560 &
9.&50 & 0.&02149 & 0.&1225 \\
0.&30 & 1.&217 &   3.&345 &
4.&00 & 0.&08717 & 0.&3879 &
10.&0 & 0.&01965 & 0.&1132 \\
\hline
\end{tabular}
\end{center}
\end{minipage}
\end{center}
\end{table*}

We improve this approximation by using the results from a
first Born approximation, and a different parameterization
has been employed altogether, namely
\begin{equation}
\label{henckenparam} {\bar n}_e(\rho) = (Z\alpha)^4
\,\left[A(x)\, L-B(x)\right]
\end{equation}
(for a list of numerical values, see Table~\ref{table2}).
There is an obvious connection between the
above two sets of functions in the limit of low nuclear
charge numbers,
\begin{equation}
F(x, Z \to 0) = A(x)\,,\quad
G(x, Z \to 0) = \frac{B(x)}{A(x)}\,.
\end{equation}
First of all, it is reassuring to verify, based on
numerical data presented in
Refs.~\cite{LeMi2007,HeTaBa1994}, that the equality
$F(x,0)=A(x)$ is valid in the phenomenologically important
interval $x=0.01\div 10$ with an accuracy better than 5\%.

One can now take the data for $F(x,0)$ as given in
Ref.~\cite{LeMi2007} and use a least-squares method in
order to fit $G(x,0)$ to numerical data in
Table~\ref{table2}, assuming the functional form
(\ref{henckenparam}). A least-squares fit, assuming the
dependence
\begin{equation}
G(x,0) = 1.5\,\ln{(x+a)}+b\,,
\end{equation}
gives the best estimates $a=1.4$ and $b=1.9$, where the
prefactor of the logarithm is fixed by the ratio of the two
quantities discussed in Eq.~(\ref{bethemaximon}). The
approximation thus obtained differs from ${B(x)}/{A(x)}$ in
the phenomenologically most important interval $x=0.02\div
5$ by less than $2$\%.

Based on the deviation of $F(x, Z=0)$ from $F(x, Z)$ by
no more than 25\% due to Coulomb corrections for heavy
nuclei~\cite{LeMi2007}, we expect that the same deviation
is valid at comparison of $G(x, Z=0)$ and $G(x, Z)$. Taking
into account that the function $G(x, Z)$ is the subleading
term of the relative order of $1/L$, we can conclude that
the approximate expression
 \begin{eqnarray}
 \label{mainapprox}
{\bar n}_e(\rho) &\approx & (Z\alpha)^4 \,F(x,Z)\,
\left[L-1.5 \, \ln(x+1.4) - 1.9\right]\,,
  \nonumber\\
L &=& \ln{(\gamma^2)}\,, \quad x = m \, \rho\,,
 \end{eqnarray}
which involves the function $F(x,Z)$ from Ref.~\cite{LeMi2007}, has
an accuracy of the order of $5$ \%. In the calculations reported below,
we use this very expression.

%
%
\section{Calculation of Unitarity Corrections}
\label{calc}

%
%
\subsection{Unitarity corrections for the
$\maybebm{e^+e^-}$ pair production}
\label{calcee}

Unitarity corrections for the process (\ref{eepair}) have
been considered in Refs.~\cite{LeMiSe2002,BaGeKuNi2002}.
Based on Eq.~(\ref{unit2ee}), we find the unitarity
correction is
\begin{equation}
\sigma^{\rm unit}_{e^+ e^-} =
- \int_{\rho_{\min}}^{\rho_{\max}}\,
\left[1- \ee^{- \bar{n}_e(\rho)}\right] \,
\bar{n}_e(\rho)\,\dd^2{\rho}\,,
\label{unitcor}
\end{equation}
where the integration limits (minimum and maximum impact
parameters) are to be specified below. Since we are
interested in so-called ``silent events'' without any ``touching''
of the nuclei, the physically allowed minimal value of the
impact parameter is
%
\begin{equation}
\rho_{\min}=2R\,,
\end{equation}
where $R$ is the nuclear radius. {\em A priori}, the upper limit is
$\rho_{\max}=\infty$, but due to fast convergence of the integral we can
use as well $\rho_{\max}=100/m$
for a quantitative estimate of the unitarity correction.
Indeed, for large $\rho$, the asymptotics (\ref{bethemaximon}) allow
an expansion of the exponential for the vacuum persistence amplitude in
(\ref{unit1gammaD}) and in analogous expressions used for the
unitarity corrections in this article, and thus there is a sufficiently
large negative power of $\rho$ characterizing all integrands for
large $\rho$; we can thus neglect of excessively large
impact parameters in the evaluation of all unitarity corrections.
As a result, the integration region in the variable $x$
can be safely chosen as
\begin{equation}
x_0 \leq x = m  \rho \leq \xmax = 100,\;\; x_0 = m
\rho_{\min} = 2 m R.
 \label{x0}
\end{equation}
Here, $m$ is the electron mass.
Using the relevant physical parameters, we find
$x_0=0.0361$ for Au, $x_0=0.0368$ for Pb, $x_0=0.0385$ for
U and $x_0=0.0213$ for Ca. Finally, the relative magnitude
of the unitarity corrections for the considered $e^+ e^-$
pair production process (\ref{eepair}) is
\begin{equation}
\delta_{e^+e^-} =
\frac{\sigma^{\rm unit}_{e^+ e^-}}{\sigma_{\rm {Born}}}\,,
\label{unitcorrel}
\end{equation}
where the known Born cross section reads~\cite{LaLi1934,Ra1937}
\begin{equation}
\sigma_{\rm {Born}}
= \frac{28}{27\pi} \,
\sigma_0 \, \left[ L^3 - 2.198\,L^2 + 3.821\,L - 1.632 \right]
\end{equation}
with
\begin{equation}
\sigma_0=\frac{(Z_1 \alpha \, Z_2 \alpha)^2}{m^2}\,,
\qquad
L = \ln(\gamma_1\gamma_2)\,.
\end{equation}

For light nuclei with not excessive nuclear charge
number [$(Z\alpha)^4 \, L\ll 1$],
it is possible to calculate in the Born
approximation the following integrals
\begin{subequations}
\label{defCDE}
\begin{align}
\label{Cn}
C_n =& \; \frac{2\pi}{n!} \,
\int_{x_0}^{\xmax} F^n(x) \, x \, \dd x\,,\qquad n=2,3,4,
\\[2ex]
\label{D2}
D_2 =& \; 2\pi \int_{x_0}^{\xmax} F^2(x)\, G(x) \, x \,\dd x\,,
\\[2ex]
\label{E2}
E_2 =& \; \frac{2\pi}{2!} \,
\int_{x_0}^{\xmax} F^2(x) \, G^2(x) \, x \, \dd x\,,
\end{align}
\end{subequations}
where $F(x)\equiv F(x,Z=0)$ and $G(x)\equiv G(x,Z=0)$. They
can be use to calculate, for light nuclei,
\begin{itemize}
\item the unitarity correction for light nuclei,
\begin{subequations}
\label{resunitlight}
\begin{equation}
\label{resunitlight1} \sigma_{e^+e^-}^{\rm unit} = -2
\frac{(Z\alpha)^8}{m^2} \, \left( C_2 \, L^2 - D_2 \, L +
E_2\right)\,,
\end{equation}
\item the total cross section for the production of
two $e^+e^-$ pairs in collisions of light nuclei,
\begin{equation}
\label{resunitlight2}
\sigma_2 = \frac{(Z\alpha)^8}{m^2} \,
\left( C_2 \, L^2 - D_2 \, L + E_2\right)\,,
\end{equation}
\item and the leading logarithmic
asymptotics for the total cross section
$\sigma_n$ for $n$-pair production with $n > 2$,
\begin{equation}
\label{resunitlight3}
\sigma_n = \frac{(Z\alpha)^{4n}}{m^2}\,C_n \, L^n\,.
\end{equation}
\end{subequations}
\end{itemize}

Using the parameterization (\ref{mainapprox}), we obtain
the following numerical results for the coefficients listed
in Eq.~(\ref{defCDE}), which enter the formulas listed in
Eqs.~(\ref{resunitlight1}), (\ref{resunitlight2}), and
(\ref{resunitlight3}) for various unitarity corrections and
cross sections [in all integrals we used the integration
region (\ref{x0}) with $x_0=0.0213$, which is the value
obtained for a typical nucleus of low charge number, namely
Ca]
 \begin{eqnarray}
 \label{CDE}
C_2& =& 2.21\,, \quad C_3 = 0.443\,,\quad C_4 =
0.119\,,
 \\
D_2& =& 15.5\,,\quad E_2 = 28.9\,.
  \nonumber
 \end{eqnarray}
The results from the previous investigations in
Ref.~\cite{LeMiSe2002} concern only those coefficients
which can be defined exclusively in terms of $F(x)$. The
previous results read $C_2=1.33$, $C_3= 0.264$, and
$C_4=0.066$. These differ from the new results listed in
(\ref{CDE}), because in the previous investigation, a less
accurate representation of $F(x)$ was used, which leads to
discrepancies especially when higher powers of $F(x)$ enter
the integrands as given in (\ref{defCDE}).

As an example, using the result (\ref{CDE}),
we found that the cross section for production of two
$e^+e^-$ pair for the Ca-Ca collisions at the LHC collider
($\gamma=3700$) is
\begin{equation}
\sigma_2=0.114\;\;\mbox{barn}\,.
\end{equation}
For heavier nuclei, one cannot use the Born approximation
$F(x, Z) \approx F(x,Z=0) \equiv F(x)$ any more. In this
case, one has to resort to numerical data given in the
Table 1 of Ref.~\cite{LeMi2007} for the heavy collision
systems Au-Au and Pb-Pb, and employ the relativistic
factors as given in Table~\ref{table1}. Indeed, unitarity
corrections for the process $ Z_1 \, Z_2\to Z_1 \, Z_2 \,
e^+ \, e^- $ have been considered in Ref.~\cite{LeMiSe2002}
and estimated as $ \delta_{e^+e^-} = -4.1 \% \;\;\mbox{for
Au-Au at RHIC}$ and $ \delta_{e^+e^-} = -3.3 \%
\;\;\mbox{for Pb-Pb at LHC}$, where exactly the
relativistic factors as given in Table~\ref{table1} have
been employed. We recall that the ratio $\delta_{e^+ e^-}$
has been defined in Eq.~(\ref{unitcorrel}). Using
Eq.~\eqref{mainapprox}, we are now in the position to give
the new values
 \begin{eqnarray}
 \label{resunitee}
\delta_{e^+e^-} &=& -5.0 \% \;\;\mbox{for the RHIC}\,,
 \\
\delta_{e^+e^-} &=& -4.0 \% \;\;\mbox{for the LHC}\,,
 \nonumber
 \end{eqnarray}
which differ from those obtained in Ref.~\cite{LeMiSe2002}
by about 20\%.

%
%
\subsection{Unitarity corrections for the
$\maybebm{\mu^+\mu^-}$ pair production}
\label{calcmumu}

Unitarity corrections for the process (\ref{mumupair}) have
been roughly estimated in Ref.~\cite{HeKuSe2007}. Based on
the considerations leading to Eqs.~(\ref{unit2ee}) and
(\ref{unit2gammaD}), we can immediately write down the
corresponding formula as
\begin{equation}
 \label{unit2mumu}
\sigma^{\rm unit}_{\mu^+ \mu^-} = - \int \left[ 1 -
\ee^{-\bar{n}_e(\rho)}\right]\, P_{\rm B} (\rho)
\,\dd^2\rho\,,
\end{equation}
where $P_{\rm B}(\rho)$ is the probability to produce muon
pair in collisions of two nuclei at a given impact
parameter $\rho$ in the Born approximation [the Coulomb
corrections to this probability, which correspond to
multiphoton exchange of the produced $e^+e^-$ with the
nuclei, are parametrically suppressed due to the large muon
mass and can be neglected --- see Ref.~\cite{HeKuSe2007}
for detail].

For a simple calculation, we can use an expression for
$P_{\rm B} (\rho)$ given in the leading logarithmic
approximation (LLA) in Ref.~\cite{HeKuSe2007}. It reads
\begin{equation}
\label{quadratic}
P_{\rm B}(\rho) = \frac{28}{9 \, \pi^2}\,
\frac{\left(Z_1 \alpha \, Z_2\alpha\right)^2}{(\mu\rho)^2}\,
\Phi(\rho)\,.
\end{equation}
Depending on the value of $\rho$,
the function $\Phi(\rho)$ assumes two different
asymptotic forms, as shown in Ref.~\cite{HeKuSe2007},
\begin{subequations}
\begin{align}
&\Phi(\rho) = \left[4 \ln\left( \frac{\displaystyle
\gamma}{\displaystyle \mu\rho} \right) + \ln\left(
\frac{\displaystyle \rho}{\displaystyle R} \right) \right] \, 
\ln\left( \frac{\displaystyle \rho}{\displaystyle R} \right) \;; 
\quad R \ll \rho \le \frac{\gamma}{\mu}\,,
\\[2ex]
&\Phi(\rho) = \ln^2\left( \frac{\displaystyle
\gamma^2}{\displaystyle \mu^2  \rho R} \right) \;;
\quad
\frac{\displaystyle \gamma}{\displaystyle \mu} \le 
\rho \ll 
\frac{\displaystyle \gamma^2}{\displaystyle \mu^2 R}\,.
\end{align}
\label{phi}
\end{subequations}
This expression is valid for large values of $\ln(\rho/R)$,
that is correct for LHC but not for RHIC. Therefore, we
consider below the case of the LHC collider only.

Using formulae (\ref{unit2mumu})--(\ref{phi}) we obtain
 \begin{equation}
\label{resunitmumu}
\delta_{\mu^+\mu^-} = \frac{\sigma^{\rm unit}_{\mu^+ \mu^-}}%
{\sigma_{\mu^+ \mu^-}}= -49 \%\;\;\mbox{for the LHC}\,,
\end{equation}
where $\delta_{\mu^+\mu^-}$ is of course defined as the
relative magnitude of the unitarity correction in
comparison to the Born cross section $\sigma_{\mu^+
\mu^-}$, in analogy with (\ref{unitcorrel}).

The roughly tenfold increase of the unitarity correction
(\ref{resunitmumu}) for muon pair production in comparison
to (\ref{resunitee}) for electron-positron pair production
demands a qualitative explanation. Indeed, the importance
of the unitarity correction is due to the enhanced
contribution of the region of small impact parameters in
the impact-parameter dependent muon pair production
probability (\ref{quadratic}). Due to the prefactor
$1/\rho^{2}$ in (\ref{quadratic}), the unitarity correction
is logarithmically enhanced as it involves an integration
proportional to $\int \dd^2 \rho / \rho^{2}$ over the range
of impact parameter $2 \, R < \rho < 1/m$.

%
%
\subsection{Unitarity corrections for Compton--type photoproduction}
\label{calccomp}

To a good approximation, tree-level photon emission by
nuclear bremsstrahlung is described by the block Feynman
diagrams of Fig.~\ref{fig3}. Let the cross section $\dd
\sigma_{\rm br}^a$ and $\dd \sigma_{\rm br}^b$ correspond
to the diagrams of Fig.~\ref{fig3}(a) and
Fig.~\ref{fig3}(b), respectively. Roughly speaking,
diagram~(a) describes the emission of radiation by the
Compton scattering of an equivalent photon, generated by
nucleus $1$, off nucleus $2$, whereas for diagram (b), the
situation is reversed. The cross section for
photoproduction by nuclear bremsstrahlung then is obtained
as the sum
\begin{equation}
\dd \sigma_{\rm br} = \dd \sigma_{\rm br}^a + \dd
\sigma_{\rm br}^b \,,
\end{equation}
because the interference term is small and can be safely
neglected.

In the LLA, the cross section $\dd \sigma_{\rm br}^{a}$ can
be calculated using the equivalent photon approximation, in
which it is expressed as follows:
\begin{equation}
\label{br1C}
\dd \sigma_{\rm br}^a = \dd n_1 \,\,
\dd \sigma_{\rm C}(\omega,E_\gamma,E_2, Z_2) \,.
\end{equation}
Here, $\dd n_1$ is the number of equivalent photons emitted
by nucleus $1$ in the energy interval $\dd \omega$ and the
impact parameter range $\dd^2 \rho$, and $\dd \sigma_{\rm
C}(\omega,E_\gamma,E_2, Z_2)$ is the differential cross
section for the Compton scattering off nucleus $2$, for an
energy $E_\gamma$ of the emitted photon, and an energy
$E_2$ of the second nucleus of charge number $Z_2$ and mass
$M_2$. The number of equivalent photons reads
\begin{equation}
\label{equivalent2} \dd n_1 = \frac{Z_1^2 \, \alpha}{
\pi^2} \, \frac{\dd\omega }{ \omega} \, \frac{\dd^2 \rho}{
\rho^2}
\end{equation}
with the integration region
\begin{equation}
\omega_{\min} \leq \omega\lesssim \frac{\gamma_1}{\rho}\,,
\;\;\; 2R \lesssim \rho \lesssim \rho_{\max} =
\frac{\gamma_1}{\omega_{\min}}
\label{intreg}
\end{equation}
and
\begin{equation}
\omega_{\min} =
\frac{E_\gamma}{4\gamma^2_2 (1-x_\gamma)}\,.
\end{equation}

For the Compton cross section, we can use the following
well-known expression, which is valid for a nucleus
approximated by a charged point particle. This approach
gives a good approximation at least in the region of not
too energetic photons, where the nuclear structure can be
safely neglected, and it reads
 \begin{eqnarray}
&&\dd \sigma_{\rm C}(\omega, E_2, E_\gamma, Z_2) =
 \\
&&= 4\pi \, \frac{Z_2^4 \alpha^2}{M_2^2} \,
\left[(1-x_\gamma) \, \left(y - 2y^2 + 2y^3\right) + \frac{
12}{y x^2_\gamma} \right]\, \frac{\dd
E_\gamma}{E_\gamma}\,,
 \nonumber
 \end{eqnarray}
where
\begin{equation}
x_\gamma = \frac{E_\gamma}{E_2} \,, \qquad
y = \frac{\omega_{\min}}{\omega}\,.
\end{equation}
Then we integrate Eq.~(\ref{br1C}) over $\omega$ and
write the result in the form
\begin{equation}
\dd \sigma_{\rm br}^a = \dd P_a(\rho)\, \dd^2\rho\,,
\label{eq}
\end{equation}
where the differential probability $\dd P_a(\rho)$ assumes
the form
 \begin{equation}
\dd P_a(\rho) = \frac{Z_1^2\alpha}{\pi^2} \,
\frac{\sigma_{\rm T}(Z_2)}{\rho^2} \, \left(1-x_\gamma +
\frac34\, x_\gamma^2 \right)\, \frac{\dd
E_\gamma}{E_\gamma} \,, \label{30}
\end{equation}
with the Thomson cross section
\begin{equation}
\sigma_{\rm T}(Z_2) = \frac{8\pi}{3} \frac{Z_2^4
\alpha^2}{M_2^2}\,.
 \label{9}
\end{equation}
Formally the probability in Eq.~\eqref{30} is divergent if
integrated over all photon energies. However, using reasonable
upper and lower bounds for $E_\gamma$, the probability
remains small even at the minimal impact parameter.

According to the parameter region relevant for the
equivalent-photon approximation~(\ref{equivalent2}), this
expression is valid in the dominant region $2R \leq \rho
\lesssim {\gamma_1}/\omega_{\min}$. Integrating
Eq.~(\ref{eq}) over this region, we obtain
\begin{equation}
\dd \sigma_{\rm br}^a =
2 \frac{Z_1^2\alpha}{\pi} \, {\sigma_{\rm T}(Z_2)} \,
\left(1-x_\gamma + \frac34 x_\gamma^2\right)\,
L_\gamma\, \frac{\dd E_\gamma}{E_\gamma}
\end{equation}
where
\begin{equation}
\label{Lgamma}
L_\gamma= \ln\left( {\frac{\rho_{\max}}{2R}} \right)=
\ln\left( \frac{2\gamma_1 \, \gamma^2_2 \, (1-x_\gamma)}{R \, E_\gamma}
\right) \,.
\end{equation}
Now the unitarity correction $\delta_\gamma$, expressed as
a fraction of the complete nuclear bremsstrahlung cross
section, can be obtained by considering diagram (a) alone,
\begin{equation}
\delta_a=
\frac{\dd \sigma^a_{\rm unit}}{\dd \sigma_{\rm br}^a}\,,
\end{equation}
and it  can be calculated using
\begin{equation}
\dd \sigma^a_{\rm unit} = -\int \dd P_a(\rho)\, \left[ 1 -
\ee^{-\bar{n}_e(\rho)} \right]\,\dd^2{\rho} \,.
 \label{53a}
\end{equation}
The main (logarithmically enhanced) contribution to $\dd
\sigma^a_{\rm unit}$ is given by the impact parameter
region $2R \leq \rho \lesssim m^{-1}$, and therefore, a
simple estimate can be given as 
\begin{equation}
\label{resunitgamma1} \delta_\gamma \sim
-(Z\alpha)^4\,\frac{L}{L_\gamma} \, \ln\left( \frac{1}{2Rm}
\right) \,.
\end{equation}

A more accurate calculation is based on the direct
integration of the vacuum persistence amplitude against
pair production that involves the number of produced
electron-positron pairs according to
Eq.~(\ref{milsteinparam}), and reads
\begin{equation}
 \label{deltaacomplete}
\delta_\gamma= - \frac{1}{L_\gamma} \int_{2R}^{100/m}
\left[ 1 - \ee^{-\bar{n}_e(\rho)}
\right]\,\frac{\dd\rho}{\rho}\,,
\end{equation}
where the convergence of the integral is assured by the
asymptotics given in Eq.~(\ref{bethemaximon}) and
the upper limit of $100/m$ for $\rho$ could have been
replaced by $\infty$.
An evaluation based on Eq.~(\ref{deltaacomplete})
gives the following result for $E_\gamma=1$~GeV,
 \begin{eqnarray}
 \label{resunitgamma2}
\delta_{\gamma}&=&-19 \% \;\;\mbox{for the RHIC}\,,
 \\
\delta_{\gamma}&=&-15 \% \;\;\mbox{for the LHC} \,.
 \nonumber
 \end{eqnarray}
%

%
%
\subsection{Unitarity correction for Delbr\"uck--type photoproduction}
\label{calcdel}

Photon emission in heavy-ion collisions via virtual
Delbr\"uck scattering has recently been considered in
Refs.~\cite{GiJeSe2008plb,GiJeSe2008epjc}, where
a surprisingly large cross sections was found for this case:
 \begin{eqnarray}
\sigma_{\gamma {\rm D}}&=& 14\;\mbox{barn for the RHIC}\,,
 \\
\sigma_{\gamma {\rm D}}&=& 50\;\mbox{barn for the LHC}\,.
 \nonumber
 \end{eqnarray}
The main contribution to these cross section comes from the
photon-energy region
\begin{equation}
m\ll E_\gamma \ll \gamma m\,.
\end{equation}
Below, we estimate the unitarity correction for the
discussed process.

The probability $P_{\gamma {\rm D}}(\rho)$, entered the
cross section (\ref{unit2gammaD}) for the photoproduction
via the virtual Delbr\"uck scattering, can be easily
obtained as a function of the impact parameter $\rho$ in
the dominant range $1 \ll m\rho \ll \gamma^2$, but the
unitarity correction is mainly given by integration in the
region $m\rho\sim 1$. Therefore, while we estimate the
unitarity correction here, we stress that a more accurate
calculation would require a direct evaluation of the
Delbr\"{u}ck--type photoproduction probability $P_{\gamma
{\rm D}}(\rho)$ in the range $m\rho\sim 1$, which was
beyond our scope.

For the sake of simplicity, we consider here only the
symmetric case $\gamma = \gamma_1 = \gamma_2$ with
identical nuclei $Z = Z_1 = Z_2$. The cross section
$\sigma_{\gamma {\rm D}}$ can be expressed by the
integration of $P_{\gamma {\rm D}}(\rho)$ over the impact
parameter via the relation
\begin{equation}
\sigma_{\gamma {\rm D}} = \int  P_{\gamma {\rm
D}}(\rho)\,\dd^2{\rho}\,.
\end{equation}
In LLA  we can use the differential cross section in the
same form as in (\ref{br1C}):
\begin{equation}
\dd \sigma_{\gamma {\rm D}}  = 2\,\dd n_\gamma \;
\sigma_{\rm D}( Z_2)
\end{equation}
where $\sigma_{\rm D}(Z)$ is a high-energy limit of the
Delbr\"{u}ck scattering cross section defined according to
Eq.~(7) of Ref.~\cite{GiJeSe2008epjc} and
\begin{equation}
\label{ranges} \dd n_\gamma = \frac{Z^2\alpha}{ \pi^2}\,
\frac{\dd\omega }{ \omega}\, \frac{\dd^2 \rho}{ \rho^2} \,,
\end{equation}
is the number of the equivalent photons. The factor 2 in
right-hand-side of $\dd n_\gamma$ takes into account two
possibilities, corresponding two diagrams in
Fig.~\ref{fig4}. This expression is valid in a parameter
range satisfying the two conditions ${m}/{\gamma} \lesssim
\omega\lesssim {\gamma}/{\rho}$ and ${1}/{m} \lesssim \rho
\lesssim {\gamma^2}/{m}$. After integration (\ref{ranges})
over $\omega$, we obtain the probability $P_{\gamma {\rm
D}}(\rho)$ in the form
\begin{equation}
\label{PgammaD} P_{\gamma {\rm D}}(\rho) =
\frac{2Z^2\alpha}{\pi^2} \; \frac{\sigma_{\rm
D}(Z)}{\rho^2} \; \ln \left( {\frac{\gamma^2}{m\rho}}
\right)\,.
\end{equation}
Under the restrictions for which the approximation made in
Eq.~(\ref{ranges}) remains valid, the expression
(\ref{PgammaD}) is applicable in the dominant region
\begin{equation}
1 \ll m\rho \ll \gamma^2\,.
\end{equation}
where again the probability is small.
We rewrite (\ref{unit2gammaD}) in the form
\begin{equation}
\sigma^{\rm unit}_{\gamma {\rm D}} = - \int P_{\gamma {\rm
D}}(\rho)\, \left[1 - \ee^{-\bar{n}_e(\rho)}
\right]\,\dd^2{\rho}\,,
  \label{unitDel}
\end{equation}
and now we can use this formula in order to estimate the
relative magnitude of the unitarity correction. The
function $\bar{n}_e(\rho)$ is of the order of unity at
$\rho\sim 1/m$ and given by the expression
\begin{equation}
\bar{n}_e(\rho) \approx 0.5\,(Z\alpha)^4\,L \qquad\qquad
(\rho\approx 1/m)\,,
\end{equation}
but drops very quickly at larger impact parameters, with an
asymptotic behavior of $\bar{n}_e(\rho)\propto 1/\rho^{2}$.
Since the function $P_{\gamma {\rm D}}(\rho)$ also drops at
large impact parameters [see Eq.~(\ref{PgammaD})], the main
contribution to $\sigma^{\rm unit}_{\gamma {\rm D}}$ comes
from the region $\rho\sim 1/m$, and we can estimate the
integral (\ref{unitDel}) as follows,
 \begin{eqnarray}
\sigma^{\rm unit}_{\gamma {\rm D}} &\sim& -\int P_{\gamma
{\rm D}}(\rho) \, \bar{n}_e(\rho)\,\dd^2{\rho} \, \sim
 \\
&\sim & -P_{\gamma {\rm D}} \left(1/m \right) \,
\bar{n}_e\left(1/m\right)\, \frac{\pi}{m^2}\,.
 \nonumber
 \end{eqnarray}
Taking into account the result (\ref{PgammaD}), we obtain
the estimate
\begin{equation}
P_{\gamma {\rm D}}(\rho) \sim \frac{2Z^2\alpha}{\pi^2}\;
\sigma_{\rm D}(Z)\,m^2\,L \qquad \mbox{at} \qquad \rho \sim
1/m\,.
 \label{57}
\end{equation}
The relative magnitude of the unitarity correction thus is
\begin{equation}
\label{resunitgammaD} \delta_{\gamma {\rm D}} =
\frac{\sigma^{\rm unit}_{\gamma {\rm D}}}{\sigma_{\gamma
{\rm D}}} \sim -0.5\,(Z\alpha)^4\,.
\end{equation}
For the nuclear collisions at modern heavy-ion machines
with parameters as listed in Table~\ref{table1}, one can
estimate unitarity corrections to the photon emission
to be on the level of $-5\%$. Let us emphasize, that the form of
this correction (\ref{resunitgammaD}) is different from
that for the lepton pair production discussed in
Secs.~\ref{calcee} and~\ref{calcmumu}.

%
%
\section{Conclusions}
\label{conclu}

In this article, we have considered unitarity corrections
for $e^+e^-$ and $\mu^+ \mu^-$ production in heavy-ion
collisions, and for the production of photons by nuclear
bremsstrahlung and by virtual Delbr\"{u}ck scattering. The
main results of the current investigation can be found in
Eqs.~(\ref{resunitlight1}) for the unitarity correction to
$e^+ e^-$ production for collisions of light nuclei, in
Eq.~(\ref{resunitee}) for the same process in heavy-ion
collisions in modern colliders with parameters as
detailed in Table~\ref{table1}, for $\mu^+ \mu^-$
collisions in modern colliders [see
Eq.~(\ref{resunitmumu})], where the unitarity correction is
numerically large, and in Eqs.~(\ref{resunitgamma2}) and
(\ref{resunitgammaD}) for the unitarity correction to
photoproduction in heavy-ion machines, with allowance for
both the ordinary nuclear bremsstrahlung and the virtual
Delbr\"{u}ck scattering process.

Our results as presented for electron-positron pair
production in Eqs.~(\ref{resunitlight1})
and~(\ref{resunitee}) are based on a refined treatment of
the vacuum persistence amplitude against multi-pair
production implied by Eq.~(\ref{mainapprox}), and they
represent an update of results previously presented in
Ref.~\cite{LeMiSe2002} for the same corrections. For $\mu^+
\mu^-$ production, we update the results of
Ref.~\cite{HeKuSe2007}. For ordinary (Compton--type) and
Delbr\"{u}ck--type photoproduction, the results for the
unitarity corrections are obtained here for the first time
to the best of our knowledge.

Finally, we notice that the estimates given here for the
coefficients $C$, $D$ and $E$ in Eq.~(\ref{CDE}) also enter
the total cross section for the production of two $e^+e^-$
pairs in collisions of light nuclei [see
Eq.~(\ref{resunitlight2})] and the leading logarithmic
asymptotics for the total cross section $\sigma_n$ for
$n$-pair production with $n > 2$ [see
Eq.~(\ref{resunitlight3})]. From a phenomenological point
of view, it is important to remark that all unitarity
corrections reduce the one-photon or one-pair production
cross sections, and that they can be numerically large [see
Eqs.~(\ref{resunitmumu}) and~(\ref{resunitgamma2})].

%
%
\section*{Acknowledgments}
We are grateful to G.~Baur, V.~Fadin, I. F. Ginzburg,
A.~Milstein, N.~Nikolaev and D.~Trautmann for useful
discussions. V.G.S. acknowledges the warm hospitality of
the Institute of Theoretical Physics of Heidelberg
University and support by the Gesellschaft f\"{u}r
Schwerionenforschung (GSI Darmstadt) under contract HD--JENT. 
This work is
partially supported by the Russian Foundation for Basic
Research (code 06-02-16064) and by the Fond of Russian
Scientific School (code 1027.2008.2). U.D.J.~acknowledges
support by Deutsche Forschungsgemeinschaft (Heisenberg
program).

\section*{Appendix}

In this appendix we briefly recall some details
regarding the derivation of
Eq.~\eqref{bethemaximon}. The functions $F$ and $G$ from
this equation enter the cross section of the
process~\eqref{eepair} as follows,
\begin{equation}
\dd\sigma_{1} = {\bar n}_e(\rho)\,\dd^2{\rho}=(Z_1\alpha
Z_2\alpha)^2 \, F(x, Z)  [L-G(x,Z)]\,\dd^2{\rho}\,.
\label{total}
\end{equation}
Beyond the Born contribution,
we should take into account, for this cross section,
the so-called Coulomb corrections
with a multiphoton exchange between the produced
pair and the first or second nucleus:
\begin{equation}
\dd\sigma_{1}=\dd\sigma_{\rm Born}+\dd\sigma_{\rm Coul}\,.
\label{totalsum}
\end{equation}
The Born contribution to the functions $F$ and $G$ was
considered in detail in Ref.~\cite{LeMiSe2002}. Thus, we
only need to find the Coulomb corrections which enter
Eq.~\eqref{bethemaximon} as items proportional to
$f(Z\alpha)$.

The Coulomb contribution to the total pair cross section
was calculated in~\cite{IvSchSe1999}; it can be presented
in the form
\begin{align}
\dd\sigma_{\rm Coul} =& \; 
\dd n_1\,\sigma_{\rm Coul}(\gamma_1 Z_2 \to e^+e^-Z_2)
\nonumber\\[2ex]
& + \dd n_2\,\sigma_{\rm Coul}(\gamma_2 Z_1 \to e^+e^-Z_1)\,,
\label{Coul}
\end{align}
where
\begin{equation}
\dd n_i=\frac{Z_i^2\alpha}{\pi^2}\,\frac{\dd
\omega_i}{\omega_i}\, \frac{\dd^2 \rho}{\rho^2}
\label{dn}
\end{equation}
is the number of the equivalent photons, produced by the
$i$th nucleus, and
\begin{equation}
\sigma_{\rm Coul}(\gamma Z \to e^+e^-Z)=-\frac{28}{9}\,
\frac{Z^2\alpha^3}{m^2}\, f(Z\alpha)
\label{BM}
\end{equation}
is the Coulomb correction to the total cross section of the
photoproduction $\gamma Z \to e^+e^-Z$ taken from the
well-known Bethe-Maximon formula.
Integrating Eq.~\eqref{dn} over $\omega_i$ in the main
region
\begin{equation}
\frac{m}{\gamma_{2,1}}\lesssim \omega_{1,2} \lesssim
\gamma_{1,2}\, m,
\end{equation}
we find
\begin{align}
& \dd\sigma_{\rm Coul}= \\
& -\frac{28}{9\pi^2}\, \frac{(Z_1\alpha
Z_2\alpha)^2}{(m\rho)^2}\,\left[f(Z_1\alpha)+f(Z_2\alpha)
\right]\, \ln \left( {\frac{\gamma_1 \, \gamma_2}{m \,
\rho}} \right)\,\dd^2 \rho\,.
\nonumber
\end{align}
Comparing this expression with Eqs.~\eqref{total} and
\eqref{totalsum} and assuming $Z_1=Z_2=Z$, we obtain both terms
proportional to $f(Z\alpha)$ in Eq.~\eqref{bethemaximon}.

\end{document}